\documentclass[prl,twocolumn, preprintnumbers,floatfix,nopacs,aps,notitlepage,showpacs,superscriptaddress]{revtex4}
\usepackage{latexsym}
\usepackage{epsfig}
\usepackage{amssymb}
\newcommand{\lp}{\left(}
\newcommand{\rp}{\right)}
\newcommand{\lb}{\left[}
\newcommand{\rb}{\right]}

\newcommand{\ba}{\begin{eqnarray}}
\newcommand{\ea}{\end{eqnarray}}
\newcommand{\be}{\begin{equation}}
\newcommand{\ee}{\end{equation}}

\newcommand{\bt}{\beta}

\newcommand{\la}{\lambda}

\newcommand{\Lag}{\mathcal{L}}

\newcommand{\ud}[2]{^{#1}_{\phantom{#1} #2}}

\begin{document}
\title{Screening Modifications of Gravity through Disformally Coupled Fields}
\author{Tomi S. Koivisto$^1$, David F. Mota$^1$, Miguel Zumalac\'arregui}
\affiliation{Institute for Theoretical Astrophysics, University of
  Oslo,  N-0315 Oslo, Norway}
\affiliation{Institut de Ciencies del Cosmos, Universitat de Barcelona IEEC-UB, Marti i Franques 1, E-08028 Barcelona, Spain}
\begin{abstract}
It is shown that extensions to General Relativity, which introduce a strongly coupled scalar field, can be viable if the interaction has a {\it non}-conformal form.
Such disformal coupling depends upon the gradients of the scalar field. Thus, if the field is locally static and smooth, the coupling becomes invisible in the Solar System: this is {\it the disformal screening mechanism}. 
A cosmological model is considered where the disformal coupling triggers the onset of accelerated expansion after a scaling matter era, giving a good fit to a wide range of background observational data.  Moreover, the interaction leaves signatures in the formation of large-scale structure that can be used to probe such couplings. 
\end{abstract}
\maketitle
In the standard $\Lambda$CDM model of cosmology, the universe at the present day appears to be extremely fine tuned. The energy scale of the 
$\Lambda$ component is about $10^{-30}$ times the most naive expectations of theory, and this $\Lambda$ has begun dominating the universal energy budget at a redshift which, compared to the redshift to the formation of simplest elements, is a fraction of about $10^{-11}$. In attempts to avoid these fine tuning problems, $\Lambda$ is often generalized to a dynamical scalar field,  whose time evolution could more naturally result in the observed energy density today  \cite{Copeland:2006wr}. 

High energy physics generically predicts an interaction between scalar degree of freedom and other forms of matter, which in turn could help to explain why the field becomes dynamically important at the present epoch. 
There are myriad variations of such models, but in all of them the coupling can be effectively described by a field-dependent mass of the dark matter particle. Those Yukawa-type couplings can be motivated by a conformal relation to scalar-tensor theories, which includes also the $f(R)$ class of modified gravity \cite{Clifton:2011jh}.

However, for {\it any} other type of gravity modification, the relation between the matter and gravitational metric will be non-conformal. 
This can also be motivated e.g. in a DBI type scenario where matter is allowed to enter the additional dimensions \cite{deRham:2010eu}. 
When given by a scalar field $\phi$, the  disformal relation can be parametrized as
\begin{equation} \label{metrics}
\bar{g}_{\mu\nu} = C(\phi)g_{\mu\nu} + D(\phi)\phi_{,\mu}\phi_{,\nu}\,,
\end{equation}
where commas denote partial derivatives.
Considering the most general physical case, Bekenstein \cite{Bekenstein:1992pj} argued that both functions $C$ and $D$ may also depend upon $(\partial\phi)^2$, but we will focus on the simpler case here. Previous applications of such a relation to cosmology include varying speed of light theories \cite{Magueijo:2003gj}, inflation \cite{Kaloper:2003yf}, dark energy \cite{Koivisto:2008ak,Zumalacarregui:2010wj}, gravitational alternatives to \cite{Bekenstein:2004ne,Milgrom:2009gv} and 
extensions of \cite{Bettoni:2011fs} dark matter. The disformal generalization of coupled quintessence here introduced is a simple set-up that is useful to study generic features of the relation (\ref{metrics}) in different scenarios.

\paragraph{The disformal matter coupling.}
Our aim is to explore the novel features from the disformal coupling in the minimal setting where gravity is Einstein's and
the scalar field is coupled to a single matter species.
The two metrics enter into the action for gravity and the coupled scalar-matter system in mutually exclusive sectors
\begin{equation}
S=\int d^4x \lb \sqrt{-g}\lp \frac{1}{16\pi G}R 
+ \mathcal{L}_{\phi}\rp + \sqrt{-\bar{g}}\bar{\mathcal{L}}_{m}\rb\,,
\end{equation}
where the matter Lagrangian is constructed using $\bar g_{\mu\nu}$ from Eq. (\ref{metrics}).
The stress energy tensors definition
$$
T^{\mu\nu}_{\phi} \equiv \frac{2}{\sqrt{-g}}\frac{\delta\lp\sqrt{-g}\mathcal{L}_{\phi}\rp}{\delta g_{\mu\nu}}\,, \quad 
T^{\mu\nu}_{m} \equiv \frac{2}{\sqrt{-g}}\frac{\delta\lp\sqrt{-\bar{g}}\bar{\mathcal{L}}_{m}\rp}{\delta g_{\mu\nu}}
\,,
$$
ensures that Einstein field equations have the usual form $G^{\mu\nu}=8\pi G T^{\mu\nu}$. However, the covariant conservation of energy momentum does not hold for the coupled components separately. Instead, we obtain that 
\ba
\label{conservation}
&&\nabla_\mu T^{\mu\nu}_m  \equiv  -Q\phi^{\, ,\nu}\,, \\
\nonumber
\text{where} \quad
Q  & = &   \frac{C'}{2C}T_m - \nabla_\la\lp\frac{D}{C} \phi_{,\mu} T^{\mu\la}_m \rp + \frac{D'}{2C} \phi{,_\mu} \phi_{,\nu} T^{\mu\nu}_m
\ea
The coupling will then generically involve second derivatives, which entail the distortion of causal structure. 

For a point particle, and taking into account the correct weight of the delta function, we have
\begin{equation} \label{proper_a}
\sqrt{-\bar{g}}\bar{\mathcal{L}}_m = -\Sigma m \sqrt{-\bar{g}_{\mu\nu}\dot{x}^\mu\dot{x}^\nu}\delta^{(4)}(x-x(\lambda))\,.
\end{equation}
From the point of view of the physical frame, the proper time the particle experiences is dilated by the conformal factor $C$. In addition, the disformal factor $D$ gives a direction-dependent effect proportional to the projection of the four-velocity along the gradient of the field:
\begin{equation} \label{proper}
\dot{\bar{x}}^2 \equiv \bar{g}_{\mu\nu}\dot{x}^\mu\dot{x}^\nu = C\dot{x}^2+D(\dot{x}\cdot \partial\phi)^2\,.
\end{equation}
Extremising the proper time of the particle along its path shows that it follows the disformal geodesics:
\begin{equation} \label{geodesic}
\ddot{x}^\mu + \bar{\Gamma}^\mu_{\alpha\beta}\dot{x}^\alpha \dot{x}^\beta = 0\,,
\end{equation}
where $\bar{\Gamma}^\mu_{\alpha\beta}$ has a rather complicated form involving second derivatives of the field \cite{Zumalacarregui:2012us}.

It is possible to derive the field equation without matter energy-momentum derivatives using projections of Eq. (\ref{conservation}) \cite{Zumalacarregui:2012us}:
\begin{equation}\label{cov-field}
\mathcal M^{\mu\nu}\nabla_\mu\nabla_\nu\phi + \frac{C}{C-2DX} \mathcal Q_{\mu\nu}T^{\mu\nu}_m + \mathcal V=0\,,
\end{equation}
\vspace{-15pt}
\begin{eqnarray}
\text{where}\;  \mathcal M^{\mu\nu} 
= \Lag_{\phi,X} g^{\mu\nu} +\Lag_{\phi,XX}\phi^{,\mu}\phi^{,\nu} - \frac{D}{C-2DX}T^{\mu\nu}_m, \nonumber \\
\mathcal Q_{\mu\nu} = \frac{C'}{2C}g_{\mu\nu} + \left(\frac{C'D}{C^2} - \frac{D'}{2C}\right) \phi_{,\mu} \phi_{,\nu},
 \; X=-\frac{1}{2}(\partial\phi)^2, \nonumber
\end{eqnarray}
and $\mathcal V = \Lag_{\phi,\phi} + 2X\Lag_{\phi,X\phi}$.  
A canonical field is assumed in the following, $\Lag_\phi = X-V$.

Equation (\ref{cov-field}) is a quasi-linear, diagonal, second order, partial differential equation. Its hyperbolic character depends on the signature of the tensor $\mathcal M^{\mu\nu}$, which involves the coupled matter energy-momentum tensor. For a perfect fluid, in coordinates comoving with it, $\mathcal M\ud{\mu}{\nu}
=\delta^\mu_\nu -\frac{D}{C-2DX}\mathop{\mathrm{diag}}(-\rho,p,p,p)$. 
Positive energy density keeps the correct sign of the time derivative term if $D>0$. 
However, a large pressure can flip the sign of the spatial derivatives coefficient, introducing an instability. The present analysis focuses on non-relativistic environments, and hence $\frac{Dp}{C-2DX}\ll1$ will be further assumed. Future work will address the effects of pressure, including the circumstances under which the stability condition can break down dynamically \cite{Zumalacarregui:2012us}.

\paragraph{An example cosmological model.} 
Let us consider an application where the field acts as quintessence and the disformal coupling is used to trigger cosmic acceleration. The Friedmann equations have the usual form
\begin{eqnarray}
 H^2 + K  &=&  \frac{8\pi G}{3}( \rho + \frac{\dot{\phi}^2}{2}+V)\,,  \nonumber \\
\dot{H}+H^2 &=&  -\frac{4\pi G }{3} ( \rho + 2\dot{\phi}^2-2V)\,, \nonumber
\end{eqnarray}
but the conservation equations for matter and the scalar field have to be computed from (\ref{conservation}), (\ref{cov-field}):
\begin{equation}
\dot{\rho}+3H\rho =  Q_0\dot{\phi} \, , \qquad \ddot{\phi}+3H\dot{\phi}+V'  =  -Q_0\,, \label{kg}
\end{equation}
were the background order coupling factor reads
\begin{equation}\label{background-coupling}
Q_0 = \frac{C'-2D(3H\dot{\phi}+V'+\frac{C'}{C}\dot{\phi}^2)+D'\dot{\phi}^2}{2\lp C+D(\rho-\dot{\phi}^2)\rp}\rho\,,
\end{equation}
after solving away the higher derivatives. In the following we restrict to flat space, $K=0$.

To study the dynamics, we specify an exponential parametrization for the disformal relation and the scalar field potential
$$
C=C_0 e^{\alpha\phi/M_p}\,, \;
D=D_0 e^{\beta(\phi-\phi_0)/M_p}\,,  \;
V=V_0 e^{-\gamma\phi/M_p} 
$$
with $M_p=(8\pi G)^{-1/2}$. Besides being motivated from some high energy scenarios, the exponential forms facilitate the choice of natural scales for the constant prefactors by shifting the zero point of the field (e.g. $D_0\sim M_p^{-4}$, $V_0\sim M_p^{4}$, $C_0$ dimensionless). Furthermore, these forms allow a convenient exploration of the phase space of the system. In addition to the previously studied fixed points \cite{Amendola:1999er,Copeland:2006wr}, we find only one new, a disformal scaling solution that is not an attractor \cite{Zumalacarregui:2012us}. 

Here we present numerical results for an example model where the relation (\ref{metrics}) is purely disformal ($C_0=1$, $\alpha=0$) and only affects the dark matter component. We include radiation and baryons to consider the full realistic universe model. In this scenario, the early evolution is as in the usual exponential quintessence model, where the slope of the potential $\gamma$ has a lower limit due to the presence of early dark energy $\Omega_{\rm ede} = 3(1+w)/\gamma^2$ \cite{Copeland:2006wr}. The new features appear when the disformal factor $D\dot\phi^2$ grows towards order one. Then the clocks that tick for dark matter, $\bar{g}_{00} = -1+D\dot{\phi}^2$, slow down and make the effective equation of state for dark matter approach minus unity asymptotically. The field also begins to freeze to avoid a singularity in the effective metric $\bar{g}_{\mu\nu}$, and the universe enters into a de Sitter stage. This natural resistance to pathology was also observed in the disformal self-coupling scenario \cite{Koivisto:2008ak,Zumalacarregui:2010wj}. 

\begin{figure}
 \begin{center}
\includegraphics[width=1\columnwidth]{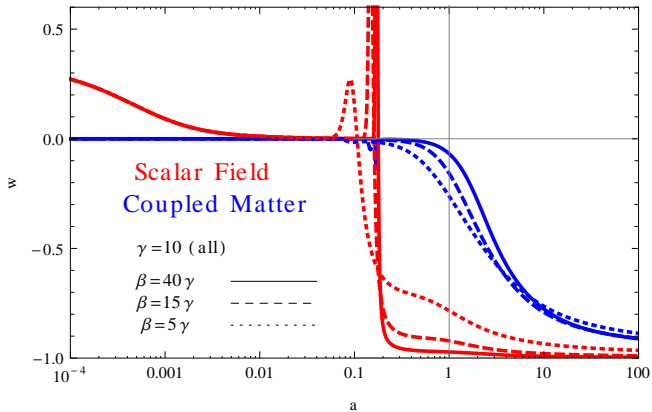}
\caption{Equation of state for the field (red) and coupled matter (blue) for different choices of the coupling slope $\beta$. High values of $\beta/\gamma$ (solid, dashed) give a good fit to observations, while low values (dotted) do not produce enough acceleration.\label{eos-fig}}
\end{center}
\end{figure}

\begin{figure}
\begin{center}
\includegraphics[width=1\columnwidth]{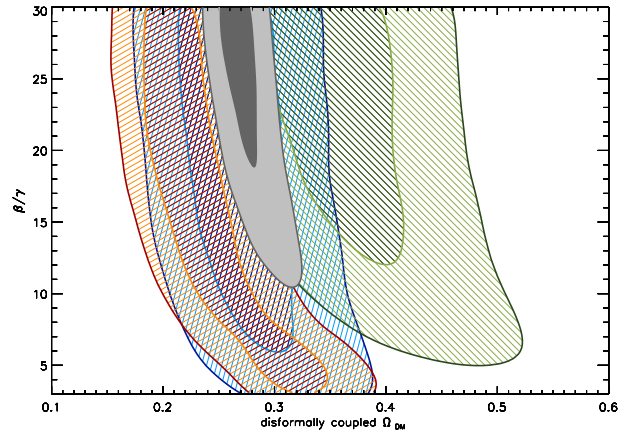}
\caption{Marginalized one and two-sigma regions obtained from Supernovae (Blue), BAO (Green), CMB angular scale + early dark energy bounds (Orange), and combined constraints. All contours included a prior on $H_0$ from the HST \cite{Riess:2011yx} and $\Omega_b H_0^2$ from Big Bang Nucleosynthesis \cite{Nakamura:2010zzi}. \label{mc-fig}}
\end{center}
\end{figure}

Thus, the disformal coupling provides a mechanism that triggers the transition to an accelerated expansion. The relatively steeper the slope of the disformal function is, i.e. the higher the ratio $\beta/\gamma$, the faster the transition happens, as seen in FIG.\ref{eos-fig}. This transition also produces a short ``bump'' in the equation of state, which may have interesting observational consequences. We performed a full background MCMC analysis with a modified version of CMBEasy \cite{Doran:2003sy} using the Union2 Supernovae compilation \cite{Amanullah:2010vv}, WiggleZ baryon acoustic scale data \cite{Blake:2011en}, cosmic microwave background angular scale \cite{Larson:2010gs} and bounds on early dark energy \cite{Reichardt:2011fv}. The obtained constraints are shown in FIG.\ref{mc-fig}. We see that for steep slopes $\gamma$ and $\beta$, the background evolution becomes increasingly similar to $\Lambda$CDM. 
At this level there are no higher bounds on these parameters, and the model is completely viable with $\chi_{\rm disf}^2=538.79$ versus $\chi^2_{\Lambda \rm CDM}=538.91$ (best fit WMAP7 parameters).
However, the model is essentially different from $\Lambda$CDM, as is quite obvious when one looks at the
effective dark matter equation of state in FIG.\ref{eos-fig}.

\paragraph{Cosmological Perturbations.} A more realistic description requires considering cosmological perturbations. In the Newtonian gauge, the linearized field equation is  
\begin{eqnarray}
\delta \ddot \phi + 3  H \delta \dot \phi + \left( \frac{k^2}{a^2} + V'' \right) \delta \phi 
 & = &  -\delta Q - 2\Phi \left(Q_0 + V'\right) \nonumber \\ & + &  \dot \phi( \dot \Phi +3 \dot \Psi ) \label{kgp} \,,
\end{eqnarray}
while the perturbed continuity and Euler equations for coupled dark matter are
\begin{eqnarray}\label{cp}
\dot \delta + \frac{\theta}{a} +\frac{Q_0}{\rho}\dot\phi \delta  
&=& 3 \dot \Psi + \frac{Q_0}{\rho} \delta\dot \phi + \frac{\delta Q}{\rho} \dot \phi \,, \\
\dot \theta + \theta\left(H + \frac{Q_0}{\rho} \dot\phi\right)  &=&  k^2 \lp\Phi + \frac{Q_0}{\rho} \delta\phi\rp \,. \label{euler}
\end{eqnarray}
The general coupling perturbation $\delta Q$ is a much more cumbersome combination of the fluid and field perturbations than in the purely conformal coupling $\delta Q^{\rm (c)} = \frac{1}{2} \log(C)'\rho\delta + \frac{1}{2}\log(C)''\rho\delta \phi$ \cite{Zumalacarregui:2012us}.

To extract the most relevant new features by analytic means, we shall consider the subhorizon approximation.
In the small scale limit, taking into account only the matter perturbations and the gradients of the field, there is a simple expression for the perturbed interaction $\delta Q$. In this Newtonian limit, we further relate the field gradient to the matter perturbation through the field equation  (\ref{kgp}), which yields the simple expression $\delta  Q^{\rm (N)} = Q_0\delta$.
Combining the dark matter equations (\ref{cp}) and (\ref{euler}) together with the usual Poisson equation, we obtain the evolution of the coupled dark matter overdensity   
\begin{equation}
\ddot{\delta}+\lb 2H+\frac{Q_0}{\rho}\dot{\phi}\rb \dot{\delta} = 4\pi G_{eff}\rho\delta\,.
\end{equation}
The Hubble friction is accompanied with an additional term due to the evolution of the field, and the source term is modulated. The last effect is captured by defining an effective gravitational constant $G_{eff}$ that determines the clustering of dark matter particles on subhorizon scales
\begin{equation} \label{d5}
\frac{G_{eff}}{G}-1=\frac{Q_0^2}{4\pi G\rho^2}\,. 
\end{equation}
This approximation has the same expression as the simple conformal case, although with a significantly different functional form of the coupling (\ref{background-coupling}). 

For our example model, the late time dependence during dark energy domination produces a large enhancement of the matter growth, 
$\delta G_{eff}/G \sim \lp \gamma V/\rho\rp^2 \gg 1$, as  $\gamma\gtrsim 10$ is required to avoid the effects of early dark energy.
Such behavior is in tension with large scale structure observations, and also occurs in conformally coupled models that attempt to address the coincidence problem \cite{Koivisto:2005nr}. This discrepancy can be alleviated with a different choice of the functions $C,D,V$. One possibility is to introduce a modulation $D(\phi)\to f(\phi)D(\phi)$ to make $Q_0$ small enough after the field enters the slow roll phase. This modification can render $\delta G_{eff}$ arbitrarily small, except for a relatively short time around the transition (see FIG.\ref{eos-fig}). A different solution to control perturbations can be achieved by allowing the field itself to live in a disformal metric, as in the uncoupled models studied in Refs. \cite{Koivisto:2008ak,Zumalacarregui:2010wj}.
Viable variations might be exploited to alleviate the claimed problems of $\Lambda$CDM with small scale structure formation \cite{Perivolaropoulos:2011hp}.
%

\paragraph{The disformal screening mechanism.} Finally, we consider the possibility of extending the disformal coupling to visible matter.
Due to the stringent bounds on equivalence principle violations \cite{Will:2005va}, some sort of screening mechanism is necessary to hide the coupling in dense environments such as the Solar System. The effects of disformal couplings to the chameleon screening were recently investigated by Noller \cite{Noller:2012sv}, who correctly noted that the disformal contribution to the conservation equations vanishes for static, pressureless configurations. This is obvious from (\ref{conservation}), since only the $T^{00}$ component is nonzero for dust, and when contracting the field derivatives with the stress tensor a non-vanishing result requires time evolution of the scalar field. 

Addressing the effects of disformal couplings therefore requires studying the field dynamics in high density, non-relativistic environments.
This regime can be explored using the general scalar field equation (\ref{cov-field}) for a stationary density distribution $\rho(\vec x)$ in the limit $\rho\to\infty$,
and neglecting the remaining spacetime curvature \footnote{More precisely, neglecting $p/\rho$, $(C-2DX)/D\rho$, $\Gamma^\mu_{00}\phi_{,\mu}$, $(C-2DX)\mathcal V/(D\rho)$, $p\phi_{,i}^2/(\rho\dot\phi^2)$ and assuming ${Dp}\ll{C-2DX}$.}. 
The same result follows from taking the limit $\rho\gg C/D,\dot\phi^2$ in (\ref{kg}):
\begin{equation} \label{time}
\ddot{\phi} \approx -\frac{D'}{2D}\dot{\phi}^2+C'\lp\frac{\dot{\phi}^2}{C}-\frac{1}{2D}\rp 
= -\frac{\bt}{2M_p}\dot{\phi^2}\,,
\end{equation}
where the first equality is general and the second applies to our example model.
The above expression departs substantially from the simple conformal coupling, for which the $\rho\to\infty$ limit is ill-defined. 
Spatial derivatives become irrelevant, as they are suppressed by a $p/\rho$ factor w.r.t. time derivatives. More importantly, the above equation becomes {\it independent of the local energy density}, making the field evolution insensitive to the presence of massive bodies. As the field rolls homogeneously, spatial gradients between separate objects, which would give rise to the scalar force, do not form.

In the purely disformal case with exponential $D$, equation (\ref{time}) can be integrated directly
\begin{equation}\label{time2}
 \dot\phi(t)=\frac{M_p}{\beta(t + t_\infty)}\,, 
\;\;\; t_\infty \equiv \frac{M_p}{\beta\dot\phi(0)}\,.
\end{equation}
In this solution, the field time variation is approximately constant while $t \ll t_\infty\propto\dot\phi(0)^{-1}$ and slows down afterwards as $\propto 1/t$. Since the coupling is proportional to $\dot\phi$, stronger couplings decay earlier. 
Furthermore, as $V$ is a decreasing function of the field, the interaction  density $\dot\rho=-(\ddot \phi + V')\dot\phi$ has at most cosmological values and its effect on dense systems is highly suppressed. Assuming $\dot\phi^2 \sim V \sim \rho_0$, $D\rho_0 \gtrsim 1$,  where $\rho_0$ is the average cosmic density, typical mass increase rates $\dot M/M$ are as small as $\sim 10^{-6}/{\rm Gy}$ for the interstellar medium and $\sim 10^{-29}/{\rm Gy}$ for the average Earth density. The induced mass variation would have consequences e.g. in the Earth-Moon system, altering the orbits in a way degenerate with a time evolution of Newton's constant. However, the predicted rates are way below current lunar laser ranging bounds $\dot G/G < 10^{-3}/{\rm Gy}$ \cite{Williams:2004qba}.  

New, potential signatures may be found in the presence of matter velocity flows, non negligible pressure, field gradients of cosmological origin, or in the mildly non-linear regime, when the screening starts taking place \cite{Zumalacarregui:2012us}. Other proposed tests involve high-precision, low-energy photon experiments \cite{Brax:2012ie} and effects on the chemical potential of the baryon-photon fluid \cite{vandeBruck:2012vq}.

This provides a novel {\it disformal screening mechanism}, which is distinct from the Vainhstein \cite{Vainshtein:1972sx}, chameleon \cite{Khoury:2003aq} and symmetron \cite{Hinterbichler:2010es} mechanisms, which respectively rely on the nonlinearity of the field derivatives and the dependence of the field mass and coupling with the ambient density.
Our mechanism relies on the existence of a well defined $\rho\to\infty$, non-relativistic limit in the scalar field equation, given by equation (\ref{time}), for which the field evolution is independent of the matter energy density. If the conformal part $C$ is negligible, only a friction term remains and the field coupling density (\ref{conservation}) is a decreasing function of time. As it evolves below its cosmological value (provided $V'<0$ and $D'/D>0$), the effects of the coupling are suppressed by a factor $\sim \rho_0/\rho$ and the theory is consistent with precision gravity tests.

In summary, the idea of disformal coupling survives scrutiny with so high flying colours that the reader must be  suspicious too. The theory has a rich structure and many interesting applications.
At cosmological scales, it can mimic closely the $\Lambda$CDM evolution, providing means to address the coincidence and cosmological constant problems. At cluster and galactic scales, new effects are to be expected due to the novel manifestation of the fifth force that could help to cure the discrepancies that the standard model has with large scale structure observations. Finally the disformal screening mechanism operating on very small scales opens a completely new avenue for fundamental scalar fields strongly coupled to matter into our reality.  
\acknowledgments{We thank L. Amendola and J. Garc\'ia-Bellido for discussions. TK and DFM are supported by the Norwegian research council. MZ is supported by MICINN through BES-2008-009090.}
\bibliography{dcm}

\begin{thebibliography}{30}
\expandafter\ifx\csname natexlab\endcsname\relax\def\natexlab#1{#1}\fi
\expandafter\ifx\csname bibnamefont\endcsname\relax
  \def\bibnamefont#1{#1}\fi
\expandafter\ifx\csname bibfnamefont\endcsname\relax
  \def\bibfnamefont#1{#1}\fi
\expandafter\ifx\csname citenamefont\endcsname\relax
  \def\citenamefont#1{#1}\fi
\expandafter\ifx\csname url\endcsname\relax
  \def\url#1{\texttt{#1}}\fi
\expandafter\ifx\csname urlprefix\endcsname\relax\def\urlprefix{URL }\fi
\providecommand{\bibinfo}[2]{#2}
\providecommand{\eprint}[2][]{\url{#2}}

\bibitem[{\citenamefont{Copeland et~al.}(2006)\citenamefont{Copeland, Sami, and
  Tsujikawa}}]{Copeland:2006wr}
\bibinfo{author}{\bibfnamefont{E.~J.} \bibnamefont{Copeland}},
  \bibinfo{author}{\bibfnamefont{M.}~\bibnamefont{Sami}}, \bibnamefont{and}
  \bibinfo{author}{\bibfnamefont{S.}~\bibnamefont{Tsujikawa}},
  \bibinfo{journal}{Int.J.Mod.Phys.} \textbf{\bibinfo{volume}{D15}},
  \bibinfo{pages}{1753} (\bibinfo{year}{2006}), \eprint{hep-th/0603057}.

\bibitem[{\citenamefont{Clifton et~al.}(2012)\citenamefont{Clifton, Ferreira,
  Padilla, and Skordis}}]{Clifton:2011jh}
\bibinfo{author}{\bibfnamefont{T.}~\bibnamefont{Clifton}},
  \bibinfo{author}{\bibfnamefont{P.~G.} \bibnamefont{Ferreira}},
  \bibinfo{author}{\bibfnamefont{A.}~\bibnamefont{Padilla}}, \bibnamefont{and}
  \bibinfo{author}{\bibfnamefont{C.}~\bibnamefont{Skordis}},
  \bibinfo{journal}{Phys.Rept.} \textbf{\bibinfo{volume}{513}},
  \bibinfo{pages}{1} (\bibinfo{year}{2012}).

\bibitem[{\citenamefont{de~Rham and Tolley}(2010)}]{deRham:2010eu}
\bibinfo{author}{\bibfnamefont{C.}~\bibnamefont{de~Rham}} \bibnamefont{and}
  \bibinfo{author}{\bibfnamefont{A.~J.} \bibnamefont{Tolley}},
  \bibinfo{journal}{JCAP} \textbf{\bibinfo{volume}{1005}}, \bibinfo{pages}{015}
  (\bibinfo{year}{2010}), \eprint{1003.5917}.

\bibitem[{\citenamefont{Bekenstein}(1993)}]{Bekenstein:1992pj}
\bibinfo{author}{\bibfnamefont{J.~D.} \bibnamefont{Bekenstein}},
  \bibinfo{journal}{Phys.Rev.} \textbf{\bibinfo{volume}{D48}},
  \bibinfo{pages}{3641} (\bibinfo{year}{1993}).

\bibitem[{\citenamefont{Magueijo}(2003)}]{Magueijo:2003gj}
\bibinfo{author}{\bibfnamefont{J.}~\bibnamefont{Magueijo}},
  \bibinfo{journal}{Rept.Prog.Phys.} \textbf{\bibinfo{volume}{66}},
  \bibinfo{pages}{2025} (\bibinfo{year}{2003}).

\bibitem[{\citenamefont{Kaloper}(2004)}]{Kaloper:2003yf}
\bibinfo{author}{\bibfnamefont{N.}~\bibnamefont{Kaloper}},
  \bibinfo{journal}{Phys.Lett.} \textbf{\bibinfo{volume}{B583}},
  \bibinfo{pages}{1} (\bibinfo{year}{2004}), \eprint{hep-ph/0312002}.

\bibitem[{\citenamefont{Koivisto}(2008)}]{Koivisto:2008ak}
\bibinfo{author}{\bibfnamefont{T.~S.} \bibnamefont{Koivisto}}
  (\bibinfo{year}{2008}), \eprint{0811.1957}.

\bibitem[{\citenamefont{Zumalacarregui
  et~al.}(2010)\citenamefont{Zumalacarregui, Koivisto, Mota, and
  Ruiz-Lapuente}}]{Zumalacarregui:2010wj}
\bibinfo{author}{\bibfnamefont{M.}~\bibnamefont{Zumalacarregui}},
  \bibinfo{author}{\bibfnamefont{T.}~\bibnamefont{Koivisto}},
  \bibinfo{author}{\bibfnamefont{D.}~\bibnamefont{Mota}}, \bibnamefont{and}
  \bibinfo{author}{\bibfnamefont{P.}~\bibnamefont{Ruiz-Lapuente}},
  \bibinfo{journal}{JCAP} \textbf{\bibinfo{volume}{1005}}, \bibinfo{pages}{038}
  (\bibinfo{year}{2010}), \eprint{1004.2684}.

\bibitem[{\citenamefont{Bekenstein}(2004)}]{Bekenstein:2004ne}
\bibinfo{author}{\bibfnamefont{J.~D.} \bibnamefont{Bekenstein}},
  \bibinfo{journal}{Phys.Rev.} \textbf{\bibinfo{volume}{D70}},
  \bibinfo{pages}{083509} (\bibinfo{year}{2004}).

\bibitem[{\citenamefont{Milgrom}(2009)}]{Milgrom:2009gv}
\bibinfo{author}{\bibfnamefont{M.}~\bibnamefont{Milgrom}},
  \bibinfo{journal}{Phys.Rev.} \textbf{\bibinfo{volume}{D80}},
  \bibinfo{pages}{123536} (\bibinfo{year}{2009}), \eprint{0912.0790}.

\bibitem[{\citenamefont{Bettoni et~al.}(2011)\citenamefont{Bettoni, Liberati,
  and Sindoni}}]{Bettoni:2011fs}
\bibinfo{author}{\bibfnamefont{D.}~\bibnamefont{Bettoni}},
  \bibinfo{author}{\bibfnamefont{S.}~\bibnamefont{Liberati}}, \bibnamefont{and}
  \bibinfo{author}{\bibfnamefont{L.}~\bibnamefont{Sindoni}},
  \bibinfo{journal}{JCAP} \textbf{\bibinfo{volume}{1111}}, \bibinfo{pages}{007}
  (\bibinfo{year}{2011}), \eprint{1108.1728}.

\bibitem[{\citenamefont{Zumalacarregui
  et~al.}(2012)\citenamefont{Zumalacarregui, Koivisto, and
  Mota}}]{Zumalacarregui:2012us}
\bibinfo{author}{\bibfnamefont{M.}~\bibnamefont{Zumalacarregui}},
  \bibinfo{author}{\bibfnamefont{T.~S.} \bibnamefont{Koivisto}},
  \bibnamefont{and} \bibinfo{author}{\bibfnamefont{D.~F.} \bibnamefont{Mota}}
  (\bibinfo{year}{2012}), \bibinfo{note}{see also suplemental material},
  \eprint{1210.8016}.

\bibitem[{\citenamefont{Amendola}(2000)}]{Amendola:1999er}
\bibinfo{author}{\bibfnamefont{L.}~\bibnamefont{Amendola}},
  \bibinfo{journal}{Phys.Rev.} \textbf{\bibinfo{volume}{D62}},
  \bibinfo{pages}{043511} (\bibinfo{year}{2000}).

\bibitem[{\citenamefont{Doran}(2005)}]{Doran:2003sy}
\bibinfo{author}{\bibfnamefont{M.}~\bibnamefont{Doran}},
  \bibinfo{journal}{JCAP} \textbf{\bibinfo{volume}{0510}}, \bibinfo{pages}{011}
  (\bibinfo{year}{2005}).

\bibitem[{\citenamefont{Amanullah et~al.}(2010)\citenamefont{Amanullah, Lidman,
  Rubin, Aldering, Astier et~al.}}]{Amanullah:2010vv}
\bibinfo{author}{\bibfnamefont{R.}~\bibnamefont{Amanullah}},
  \bibinfo{author}{\bibfnamefont{C.}~\bibnamefont{Lidman}},
  \bibinfo{author}{\bibfnamefont{D.}~\bibnamefont{Rubin}},
  \bibinfo{author}{\bibfnamefont{G.}~\bibnamefont{Aldering}},
  \bibinfo{author}{\bibfnamefont{P.}~\bibnamefont{Astier}},
  \bibnamefont{et~al.}, \bibinfo{journal}{Astrophys.J.}
  \textbf{\bibinfo{volume}{716}}, \bibinfo{pages}{712} (\bibinfo{year}{2010}),
  \eprint{1004.1711}.

\bibitem[{\citenamefont{Blake et~al.}(2011)\citenamefont{Blake, Kazin, Beutler,
  Davis, Parkinson et~al.}}]{Blake:2011en}
\bibinfo{author}{\bibfnamefont{C.}~\bibnamefont{Blake}},
  \bibinfo{author}{\bibfnamefont{E.}~\bibnamefont{Kazin}},
  \bibinfo{author}{\bibfnamefont{F.}~\bibnamefont{Beutler}},
  \bibinfo{author}{\bibfnamefont{T.}~\bibnamefont{Davis}},
  \bibinfo{author}{\bibfnamefont{D.}~\bibnamefont{Parkinson}},
  \bibnamefont{et~al.}, \bibinfo{journal}{Mon.Not.Roy.Astron.Soc.}
  \textbf{\bibinfo{volume}{418}}, \bibinfo{pages}{1707} (\bibinfo{year}{2011}).

\bibitem[{\citenamefont{Larson et~al.}(2011)\citenamefont{Larson, Dunkley,
  Hinshaw, Komatsu, Nolta et~al.}}]{Larson:2010gs}
\bibinfo{author}{\bibfnamefont{D.}~\bibnamefont{Larson}},
  \bibinfo{author}{\bibfnamefont{J.}~\bibnamefont{Dunkley}},
  \bibinfo{author}{\bibfnamefont{G.}~\bibnamefont{Hinshaw}},
  \bibinfo{author}{\bibfnamefont{E.}~\bibnamefont{Komatsu}},
  \bibinfo{author}{\bibfnamefont{M.}~\bibnamefont{Nolta}},
  \bibnamefont{et~al.}, \bibinfo{journal}{Astrophys.J.Suppl.}
  \textbf{\bibinfo{volume}{192}}, \bibinfo{pages}{16} (\bibinfo{year}{2011}).

\bibitem[{\citenamefont{Reichardt et~al.}(2012)\citenamefont{Reichardt,
  de~Putter, Zahn, and Hou}}]{Reichardt:2011fv}
\bibinfo{author}{\bibfnamefont{C.~L.} \bibnamefont{Reichardt}},
  \bibinfo{author}{\bibfnamefont{R.}~\bibnamefont{de~Putter}},
  \bibinfo{author}{\bibfnamefont{O.}~\bibnamefont{Zahn}}, \bibnamefont{and}
  \bibinfo{author}{\bibfnamefont{Z.}~\bibnamefont{Hou}},
  \bibinfo{journal}{Astrophys.J.} \textbf{\bibinfo{volume}{749}},
  \bibinfo{pages}{L9} (\bibinfo{year}{2012}), \eprint{1110.5328}.

\bibitem[{\citenamefont{Koivisto}(2005)}]{Koivisto:2005nr}
\bibinfo{author}{\bibfnamefont{T.}~\bibnamefont{Koivisto}},
  \bibinfo{journal}{Phys.Rev.} \textbf{\bibinfo{volume}{D72}},
  \bibinfo{pages}{043516} (\bibinfo{year}{2005}).

\bibitem[{\citenamefont{Perivolaropoulos}(2011)}]{Perivolaropoulos:2011hp}
\bibinfo{author}{\bibfnamefont{L.}~\bibnamefont{Perivolaropoulos}}
  (\bibinfo{year}{2011}), \eprint{1104.0539}.

\bibitem[{\citenamefont{Riess et~al.}(2011)\citenamefont{Riess, Macri,
  Casertano, Lampeitl, Ferguson et~al.}}]{Riess:2011yx}
\bibinfo{author}{\bibfnamefont{A.~G.} \bibnamefont{Riess}},
  \bibinfo{author}{\bibfnamefont{L.}~\bibnamefont{Macri}},
  \bibinfo{author}{\bibfnamefont{S.}~\bibnamefont{Casertano}},
  \bibinfo{author}{\bibfnamefont{H.}~\bibnamefont{Lampeitl}},
  \bibinfo{author}{\bibfnamefont{H.~C.} \bibnamefont{Ferguson}},
  \bibnamefont{et~al.}, \bibinfo{journal}{Astrophys.J.}
  \textbf{\bibinfo{volume}{730}}, \bibinfo{pages}{119} (\bibinfo{year}{2011}).

\bibitem[{\citenamefont{Nakamura et~al.}(2010)}]{Nakamura:2010zzi}
\bibinfo{author}{\bibfnamefont{K.}~\bibnamefont{Nakamura}} \bibnamefont{et~al.}
  (\bibinfo{collaboration}{Particle Data Group}), \bibinfo{journal}{J.Phys.G}
  \textbf{\bibinfo{volume}{G37}}, \bibinfo{pages}{075021}
  (\bibinfo{year}{2010}).

\bibitem[{\citenamefont{Will}(2005)}]{Will:2005va}
\bibinfo{author}{\bibfnamefont{C.~M.} \bibnamefont{Will}},
  \bibinfo{journal}{Living Rev.Rel.} \textbf{\bibinfo{volume}{9}},
  \bibinfo{pages}{3} (\bibinfo{year}{2005}).

\bibitem[{\citenamefont{Noller}(2012)}]{Noller:2012sv}
\bibinfo{author}{\bibfnamefont{J.}~\bibnamefont{Noller}}
  (\bibinfo{year}{2012}), \eprint{1203.6639}.

\bibitem[{\citenamefont{Williams et~al.}(2004)\citenamefont{Williams, Turyshev,
  and Boggs}}]{Williams:2004qba}
\bibinfo{author}{\bibfnamefont{J.~G.} \bibnamefont{Williams}},
  \bibinfo{author}{\bibfnamefont{S.~G.} \bibnamefont{Turyshev}},
  \bibnamefont{and} \bibinfo{author}{\bibfnamefont{D.~H.} \bibnamefont{Boggs}},
  \bibinfo{journal}{Phys.Rev.Lett.} \textbf{\bibinfo{volume}{93}},
  \bibinfo{pages}{261101} (\bibinfo{year}{2004}), \eprint{gr-qc/0411113}.

\bibitem[{\citenamefont{Brax et~al.}(2012)\citenamefont{Brax, Burrage, and
  Davis}}]{Brax:2012ie}
\bibinfo{author}{\bibfnamefont{P.}~\bibnamefont{Brax}},
  \bibinfo{author}{\bibfnamefont{C.}~\bibnamefont{Burrage}}, \bibnamefont{and}
  \bibinfo{author}{\bibfnamefont{A.-C.} \bibnamefont{Davis}}
  (\bibinfo{year}{2012}), \eprint{1206.1809}.

\bibitem[{\citenamefont{van~de Bruck and Sculthorpe}(2012)}]{vandeBruck:2012vq}
\bibinfo{author}{\bibfnamefont{C.}~\bibnamefont{van~de Bruck}}
  \bibnamefont{and}
  \bibinfo{author}{\bibfnamefont{G.}~\bibnamefont{Sculthorpe}}
  (\bibinfo{year}{2012}), \eprint{1210.2168}.

\bibitem[{\citenamefont{Vainshtein}(1972)}]{Vainshtein:1972sx}
\bibinfo{author}{\bibfnamefont{A.}~\bibnamefont{Vainshtein}},
  \bibinfo{journal}{Phys.Lett.} \textbf{\bibinfo{volume}{B39}},
  \bibinfo{pages}{393} (\bibinfo{year}{1972}).

\bibitem[{\citenamefont{Khoury and Weltman}(2004)}]{Khoury:2003aq}
\bibinfo{author}{\bibfnamefont{J.}~\bibnamefont{Khoury}} \bibnamefont{and}
  \bibinfo{author}{\bibfnamefont{A.}~\bibnamefont{Weltman}},
  \bibinfo{journal}{Phys.Rev.Lett.} \textbf{\bibinfo{volume}{93}},
  \bibinfo{pages}{171104} (\bibinfo{year}{2004}).

\bibitem[{\citenamefont{Hinterbichler and Khoury}(2010)}]{Hinterbichler:2010es}
\bibinfo{author}{\bibfnamefont{K.}~\bibnamefont{Hinterbichler}}
  \bibnamefont{and} \bibinfo{author}{\bibfnamefont{J.}~\bibnamefont{Khoury}},
  \bibinfo{journal}{Phys.Rev.Lett.} \textbf{\bibinfo{volume}{104}},
  \bibinfo{pages}{231301} (\bibinfo{year}{2010}), \eprint{1001.4525}.

\end{thebibliography}
\end{document}